\documentstyle[12pt]{article}
\begin{document}

\def\a{\alpha}
\def\b{\beta}
\def\ch{\chi}
\def\d{\delta}
\def\e{\epsilon}
\def\f{\phi}
\def\g{\gamma}
\def\h{\eta}
\def\i{\iota}
\def\j{\psi}
\def\k{\kappa}
\def\l{\lambda}
\def\m{\mu}
\def\n{\nu}
\def\o{\omega}
\def\p{\pi}
\def\q{\theta}
\def\r{\rho}
\def\s{\sigma}
\def\t{\tau}
\def\u{\upsilon}
\def\x{\xi}
\def\z{\zeta}
\def\D{\Delta}
\def\F{\Phi}
\def\G{\Gamma}
\def\J{\Psi}
\def\L{\Lambda}
\def\O{\Omega}
\def\P{\Pi}
\def\S{\Sigma}
\def\U{\Upsilon}
\def\X{\Xi} 
\def\T{\Theta}
\def\vf{\varphi}

\def\Ab{\bar{A}}
\def\gi{g^{-1}}
\def\li{{ 1 \over \l } }
\def\lb{\l^{*}}
\def\zb{\bar{z}}
\def\ub{u^{*}}
\def\vb{v^{*}}
\def\Tb{\bar{T}}
\def\pp {\partial }
\def\pb {\bar{\partial }}
\def\be{\begin{equation}}
\def\ee{\end{equation}}
\def\ben{\begin{eqnarray}}
\def\een{\end{eqnarray}} 
\def\Tr{\rm{Tr}} 
\def\ddt{\hat}
\def\dddt{\acute}

\hsize=16.5truecm
\addtolength{\topmargin}{-0.8in}
\addtolength{\textheight}{1in}
\hoffset=-.5in

\thispagestyle{empty}
\begin{flushright} \ November \ 1996\\
SNUCTP 97-079\\
\end{flushright}
\begin{center}
 {\large\bf Grassmann manifold Bosonization of QCD in Two Dimensions  }\\[.1in]
\vglue .5in
 KyoungHo Han\footnote{ E-mail address; khan@photon.kyunghee.ac.kr }
\\[.2in]
{and}
\\[.2in]
H. J. Shin\footnote{ E-mail address; hjshin@nms.kyunghee.ac.kr }
\\[.2in]
{\it  
Department of Physics \\
and \\
Research Institute of Basic Sciences \\
Kyunghee University\\
Seoul, 130-701, Korea}
\\[.2in]
{\bf ABSTRACT}\\[.2in]
\end{center}
Two dimensional QCD is bosonized to be an integrably deformed
Wess-Zumino-Witten model under proper limit. 
Fermions are identified having indices of the Grassmann manifold. 
Conditions for integrability are analyzed and their physical meanings
are discussed.
We also address the nature of the exactly solvable part of the theory 
and find the infinitely many conserved quantities.
\vglue .1in

\newpage
\section{Introduction}
Understanding QCD, for example the quark confinement, is an important 
problem in theoretical physics which is
not solved yet. One promising approach which receives much attention is
the bosonization. t'Hooft found that QCD becomes
equivalent to an effective field theory of mesons in the limit of a large
number of colors (large-M limit).\cite{hooft} 
Elaborating this idea, the Skyrme model was revised 
as an effective QCD theory where baryons are described by
the solitions of the theory.\cite{skyrme}\cite{witten} It was found that this
scheme is very helpful in describing various properties of nucleons
including their masses.\cite{adkins} However the bridge between this 
phenomenological model and the physics of basic constituents is still missing.
In this respect a lot of effort was devoted to two dimensional(2D) theory where
a low-energy effective action can be derived directly from 2D-QCD using
non-abelian bosonization.\cite{frish1}\cite{frish2}\cite{frish3}\cite{frish4}\cite{frish5}
Especillay in \cite{frish1} it was shown that the
effective low energy theory of the two dimensional QCD when bosonized 
is described by the Wess-Zumino-Witten(WZW)
model with a mass term. 
This effective theory successfully describes the baryon spectrum
and their various physical properties.\cite{frish5} 
But the integrable structure of the theory is not
known and the baryons are described by the low energy classical solutions 
and not by the
solitons.  As the integrable structure of the theory is important to understand
the stability of baryons it is strongly desired to find a bosonization scheme
with this structure.

In this paper we describe a new way of bosonization using the conformal
embedding structure of conformal field theories that was introduced in
\cite{GNO}\cite{arcuri}. This formalism leads to a bosonized theory
of 2D-QCD  which is integrable under proper limit. The integrable part of the
resulting theory is a kind of nonabelian generalization of the sine-Gordon
model\cite{shin1} and it can be thought as a bosonization of a generalized massive Thirring model with a U(1) current-current
interaction. This theory is described
by an integrably deformed WZW model having Grassmann manifold as a 
symmetric space, see Ref.\cite{shin1}. The deformation term corresponds 
to a fermion mass which only one flavor obtains. 
A similar analysis as in \cite{shin1}\cite{shin2}\cite{shin3}\cite{shin4}
can be applied to the 
present theory leading to a construction of the Lax pair,
solitons and 
Backlund transformation. The theory corresponds to the bosonization of 2D-QCD 
when we remove the U(1) current interaction term. This interaction
term which is related to the particle pair production, does not significantly change
the integrability of the theory as it
does not alter the degenerate structure of vaccum preserving the topological
notion of solitons. We can make the 
contribution of this term negligible by taking
proper limit. Main results of our analysis of 2D-QCD bosonization 
with the integrability condition are (related to 4-dimensional real world);
1)2D integrability under proper limit(baryons emerge as solitons 
in the theory of mesons), 2)only one fermion is massive in 2D theory(
top quark is
much massive than other quarks), 
3)Grassmann manifold is descibed by 
$SU(M +N)/SU(M) \times SU(N) \times U(1)$ symmetric space
(quarks have only two quantum numbers, i.e.
flavor and color).
Generalization of our method for bosonizing the massive GNO fermions 
corresponding to other symmetric spaces in Ref.\cite{GNO} will be
appeared elsewhere.\cite{park}

\section{GNO fermions and bosonization}
The action of two dimensional QCD with $M$ color and $N$ flavor
massive Dirac fermions is 
\be
S=  \int d^2 x [\sum_{a=1,i=1}^{M,N} \bar \J ^{ai} \{-i \d_a^b 
\not \! \pp -
 \not \! \! A_\a (T^\a)_a^b \} \J_{bi} +\sum_{i=1}^N m_i^{(q)} 
\sum_{a=1}^M \bar{\J}^{ai} \J_{ai}
-{1 \over 2 e_c ^2} {\Tr} F_{\m \n} F^{\m \n} ],
\ee
where $T^\a$ are the generators of the color $SU(M)$ group. Note that we assign
different fermion masses $m_i^{(q)}$ for each flavor.
This theory has $U(MN)$ global symmetry when we neglect the mass term.
The analysis of 2D-QCD through ``non-abelian bosonization" introduced by
Witten\cite{Witten} have been achieved
in \cite{frish1}\cite{frish2}\cite{frish3}\cite{frish4}\cite{frish5}.
These papers analyze 2D-QCD in the semi-classical limit and obtain the 
low-lying baryon spectrum\cite{frish1}\cite{frish2}, multibaryons\cite{frish3} 
and matrix elements like various
quark content.\cite{frish4} Their treatment was in the spirit of Skyrme
model\cite{skyrme} where baryons composed of M quarks can be treated as simple
solitons in the bosonic language. The bosonized theory of 2D-QCD is
the gauged WZW model with action on $U(MN)$ group manifold;
\ben
S[f, A, \Ab]&=&S_{WZW} [f] \nonumber \\
&+&{1 \over 2\p} \int d^2 x {\Tr} [
i A f \pb f^\dagger + i \Ab f^\dagger \pp f - A f \Ab f^\dagger + \Ab A]
\nonumber \\
&+&\sum {m_i^2 \over 2 \p} 
N_\m \int d^2 x {\Tr} (f_{ai,ai} + f_{ai,ai}^\dagger) 
-{1 \over 2 e_c ^2} \int d^2 x {\Tr} F_{\m\n} F^{\m\n}
\een 
Here $S_{WZW} [f]$ is the action of the WZW theory with $f \subset U(MN)$.
$\pp$($\pb$) denotes the derivative with respect to $z=x+t$($\zb = x-t$).
$N_\m$ denotes normal ordering at mass $\m$
and the bosonic mass $m_i$ is
related to $m_i^{(q)}$ and $\m$.\cite{gonzales} This theory, though
quite interesting, does not permit the integrability analysis which
is essential for Skyrme model approach. 

Instead of above formalism we use in this paper a gauged WZW model 
with action on 
$G = SU(M) \times SU(N) \times U(1)$ group manifold. The idea is to express 
$f \subset U(MN)$ matrix elements in eq.(2)
in terms of the group elements $g \subset G$ and makes the theory integrable.
The key expression for our approach is
\be
f_{ai,a'i'} ={1 \over 2} {\Tr} g^{-1} p_{ai}^{(1)} g (p_{a'i'}^{(1)}
-i p_{a'i'}^{(2)}) ={1 \over 2} {\Tr} g^{-1} p_{ai}^{(2)} g (p_{a'i'}^{(2)}
+i p_{a'i'}^{(1)}).
\ee
In the following we explain the motivation of our approach and notations used in eq.(3). 
The WZW model on the $G= SU(M) \times SU(N) \times U(1)$ group which is obtained using eq.(2),(3) is conformally equivalent to the original
WZW model on $F=U(MN)$ group, i.e. the difference of their two Virasoro
algebras, $\cal{L}_F -\cal{L}_G$ has vanishing c-number. This fact is
due to the underlying symmetric space structure in eq.(3).  Ref.\cite{arcuri}
indeed shows that all possible subalgebras of classical algebra which are conformally embedded to a larger group $F$
can be found directly from the
known classification of symmetric spaces. The symmetric space
$ G'/G$ for our concern is the AIII type Grassmann manifold with
$G' = SU(M+N)$. Related with this conformal embedding structure is the 
theorem due to Goddard, Nahm and Olive\cite{GNO}
(GNO for short); a necessary 
and sufficient condition for the algebraic coincidence
of Sugawara energy momentum tensor with that of the free fermions is
there exists a group $G' \subset G$ such that $G'/G$ 
is a symmetric space
with the fermions transforming under $G$ just as the tangent space to
$G'/G$ does.
Based on this theorem they found all the fermionic theories for which an
equivalent WZW bosonic action can be constructed.  
Bosonization of 2D-QCD exactly fits 
in this category with $F = U(MN)$, 
$G' = SU(M+N)$ and $G = SU(M) \times SU(N) \times U(1)$ groups.
It is interesting to note that there is no symmetric space which has $G = SU(L) \times
SU(M) \times SU(N) \times U(1)$ which means fermions at most have two
quantum nembers, flavor and color.
Let us express the group elements $g \subset G'$ which also belongs to
the group G as $(M+N) \times (M+N)$ 
matrix;
\be
g=\left(
\begin{array}{cc} g_N \subset SU(N) & 0 \\
0 & g_M \subset SU(M)
\end{array} \right) e^{\q T},
\ee
with
\be
T={-i \over M+N} \left( \begin{array}{cc}M I_{N \times N} & 0 \\
0 & -N I_{M \times M} \end{array} \right).
\ee
Type AIII space, called the hermitian symmetric
space has the complex structure allowing following decomposition.  Its $\bf p$ generators, the vector space 
complement of Lie
algebra $\bf{g}$ of $G$ in $\bf{g'}$ of $G'$, i.e. $\bf{g'} = \bf{g} \oplus \bf{p}$,
can be partitioned into MN families having two elements each;
$p_{ai} ^{(1)},\  p_{ai} ^{(2)} (\ a=1,M;\ i=1,N$).
Explicitly their matrix elements are
$(p_{ai}^{(1)})_{lm} = \d_{a+N,m} \d_{i,l} +\d_{a+N,l} \d_{i,m}
(\ l,m=1,$M+N) and
$(p_{ai}^{(2)})_{lm} = -i \d_{a+N,m} \d_{i,l} +i \d_{a+N,l} \d_{i,m}$.  
They satisfy following commutation relations
\be
[T, p_{ai} ^{(1)}]=p_{ai} ^{(2)},\ \  [T, p_{ai} ^{(2)}]=-p_{ai} ^{(1)}.
\ee 
Using properties $p_{ai} ^\dagger = p_{ai}, \ {\Tr} p_{ai} ^{(k)} p_{a'i'}
 ^{(k')} =2 \d_{aa'} \d_{ii'} \d_{kk'}$ we can easily show that
\be
g^{-1}  p_{ai} ^{(1)} g = {1 \over 2} f_{ai,a'i'} (p_{a'i'} ^{(1)} + i
p_{a'i'} ^{(2)}) + {1 \over 2} f_{ai,a'i'}^* (p_{a'i'} ^{(1)} - i
p_{a'i'} ^{(2)}).
\ee
Then the unitarity of $f$ is proved as
\be
2 \d_{aa',ii'} = {\Tr} (g^{-1} p_{ai} ^{(1)} g) (g^{-1} p_{a'i'} ^{(1)} g)
=2 (f f^\dagger)_{aa',ii'}.
\ee
The equality of first and second expression for $f$ in eq.(3) can be
shown using the relation $e^{-\q T} p_{ai}^{(1)} e^{\q T} = \cos \q p_{ai}^{(1
)} -\sin \q p_{ai}^{(2)}$ and the fact that $U(1)$ commutes with every
elements of $G$.

We now express the bosonic currents $C_{a'i',ai} ^{m(M)} (f^{-1} 
\pp f)_{ai,a'i'}$ and $C_{a'i',ai} ^{n(N)} (f^{-1} 
\pp f)_{ai,a'i'}$ in terms of the group $G$ bosonic
currents using eq.(3). Here $C_{a'i',ai} ^{m(M)}$ and $C_{a'i',ai} ^{n(N)}$
are defined as
\be
[T_m ^{(M)}, p_{ai} ^{(1)} - i p_{ai} ^{(2)} ]=
i( p_{a'i'} ^{(1)} - i p_{a'i'} ^{(2)} ) C_{a'i',ai} ^{m(M)},\ 
[T_n ^{(N)}, p_{ai} ^{(1)} - i p_{ai} ^{(2)} ]=
i( p_{a'i'} ^{(1)} - i p_{a'i'} ^{(2)} ) C_{a'i',ai} ^{n(N)},
\ee
where  $T_m ^{(M)} (m=1,M^2-1), T_n ^{(N)} (n=1,N^2-1)$ are generators of 
Lie algebra $\bf g' =su(m+n)$ corresponding to the subalgebra $\bf su(m)$ and 
$\bf su(n)$ each. They are normalized as ${\Tr} T_m ^{(M)} T_{m'} ^{(M)}
= 2 \d_{m m'}$,  ${\Tr} T_n ^{(N)} T_{n'} ^{(N)}= 2 \d_{n n'}$ and
 ${\Tr} T_m ^{(M)} T_{n} ^{(N)}=0$. Explicitly $C_{a'i',ai} ^{m(M)}=
-i \d_{ii'} (T_m ^{(M)})_{a'+N,a+N}$ and $C_{a'i',ai} ^{n(N)}=
i \d_{aa'} (T_n ^{(N)})_{i,i'}$ and they have following properties
\be
[ p_{ai} ^{(1)} - i p_{ai} ^{(2)} , p_{a'i'} ^{(1)} + i p_{a'i'} ^{(2)}]
=-2i C_{ai,a'i'} ^{m(M)} T_m ^{(M)} -2i C_{ai,a'i'} ^{n(N)} T_n ^{(N)}
+4i {M+N \over MN} \d_{aa'} \d_{ii'} T
\ee
and 
\be
C_{ai,a'i'} ^{m(M)} C_{a'i',ai} ^{m'(M)}=-2N \d_{mm'},\ \ 
C_{ai,a'i'} ^{n(N)} C_{a'i',ai} ^{n'(N)}=-2M \d_{nn'},\ \
C_{ai,a'i'} ^{m(M)} C_{a'i',ai} ^{n(N)}=0.
\ee
Using all these properties the $SU(M)$ current $C_{a'i',ai} ^{m(M)} (f^{-1} 
\pp f)_{ai,a'i'}$ becomes
\ben
 & &{1 \over 2} C_{a'i',ai} ^{m(M)} (f^{-1})_{ai,a''i''} {\Tr} g^{-1}
p_{a''i''}^{(1)} g [ g^{-1} \pp g, (p_{a'i'} ^{(1)} - i p_{a'i'} ^{(2)})]
\nonumber \\
  &=&iC_{a'i',ai} ^{m(M)} \d_{aa''} \d_{ii''} C_{a''i'',a'i'} ^{m'(M)} 
(g^{-1} \pp g)_{m'} ^{(M)}
= -2i N (g^{-1} \pp g)_m ^{(M)}
\een
where $g^{-1} \pp g =  (g^{-1} \pp g)_m ^{(M)} T_m ^{(M)}
+(g^{-1} \pp g)_n ^{(N)} T_n ^{(N)} + \pp \q T.$  Similarly the $SU(N)$ current
is $C_{a'i',ai} ^{n(N)} (f^{-1} 
\pp f)_{ai,a'i'} = -2i M (g^{-1} \pp g)_n ^{(N)}$ and $U(1)$ current is
$(f^{-1} \pp f)_{ai,ai} = i MN \pp \q.$ These expressions show that the
currents satisfy level-N $SU(M)$ and level-M $SU(N)$ Kac-Moody
algebras each, which can be understood as the bosonic
correspondent of GNO result.\cite{GNO}

The action of gauged WZW theory, eq.(2),  can also be expressed in terms of the elements
of group G. For example, the kinetic term
is calculated to be
\ben
 & &(f^{-1} \pp f)_{ai,a'i'} (f^{-1} \pb f)_{a'i',ai} \nonumber \\
  &=& 2N (g^{-1} \pp g)_m ^{(M)}
(g^{-1} \pb g)_m ^{(M)} +2M (g^{-1} \pp g)_n ^{(N)} (g^{-1} \pb g)_n ^{(N)}
-MN \pp \q \pb \q \nonumber \\
  &=&M {\Tr} g_1 ^{-1} \pp g_1 g_1 ^{-1} \pb g_1 + N {\Tr} 
g_2 ^{-1} \pp g_2 g_2 ^{-1} \pb g_2 -MN \pp \q \pb \q,
\een
where $g_1$($g_2$) means the group element $g$ in eq.(4) 
with $g_M=e ^{\q T}=1$($g_N=e ^{\q T}=1$). 
When we take the gauge field $A_{ai,a'i'}=
A_m  (T_m ^{(M)})_{a+N,a'+N} \d_{ii'}$, terms containing gauge field
are also expressed by $A = A_m  T_m ^{(M)}$ and $g$. 
In this way the gauged WZW action, eq.(2), 
of $MN \times MN$ matrix $f$
can be rewritten as an action of $(M+N) \times (M+N)$ matrix $g$;
\ben
 & &M S_{WZW}[g_1] + N S_{WZW}[g_2]  -{1 \over 2 e_c^2}F_{\m\n} F^{\m\n}
+{m_i ^2 \over 2 \p} N_\m {\Tr} 
(g_1 g_2 e^{\q T})^{-1} p_{ai} ^{(1)} (g_1 g_2 e^{\q T})
p_{ai} ^{(1)}
 \nonumber \\
 &+&  \int d^2x [ -{MN \over 8 \p} \pp \q \pb \q
+{N \over 2 \p}{\Tr}(i A g_2 \pb g_2 ^{-1} +i \Ab g_2 ^{-1} \pp g_2 
- A g_2 \Ab g_2 ^{-1} 
+ A \Ab)].
\een
To treat the gauge field we follow the prescription of Ref.\cite{frish5}.
First take the gauge $\Ab =0$ and integrate out the $A$ field to obtain
an action having $-({ e_c N \over 4 \p})^2  \int d^2 x {\Tr} H^2$. Here
$H$ is defined by $\pb H =i g_2 \pb g_2 ^{-1}$ with the boundary
condition $H(- \infty, \zb ) = 0 $. In the strong coupling limit 
$e_c / m_i \rightarrow \infty$ the fields $g_2$ which contribute to $H$ will become infinitely
heavy and can be ignored. We then get the low-energy effective action
with a substitution $g_2 =1$ and a suitable identification of the mass $m_i$.
To make an integrable theory from the QCD theory in eq.(14), we take
two modifications on the theory together with $g_2 =1$; take the mass $m_i = 0$
except $i=1$ and add an interaction term of Thirring type 
$-{\p \over 2M (N+1)} \bar{\J}_{ai} \g_+ \J_{ai} \bar{\J}_{a'i'} \g_- \J_{a'i'}$.
The action corresponding to the integrable part of the effective theory, 
which we will explain below, is
\be
S = M S[l] +M {m^2 \over 2 \p} \int d^2 x {\Tr} l^{-1} p l p
\ee
where $m$ is related to the quark mass $m_1$.\cite{gonzales} 
Here $l$,$p$ are $(N+1) \times (N+1)$ matrices 
\ben
l &=& \left( \begin{array}{cccc} &&& \\ & g_N \subset SU(N) & & 0\\ &&& \\
 & 0 & & 1 \end{array} \right) 
\left( \begin{array}{cccc}  e^{i \f} & \cdots & & 0\\
0 & e^{i \f} & & \\ & \cdots & & \\
 & 0 & & e^{-i N\f} \end{array} \right) 
 \nonumber \\
p &=& \left( \begin{array}{cccc} 0&0& \cdots &1\\ &&& \\
0 &0 & \cdots &0\\ &\cdots
\\1 & 0& \cdots &0 \end{array} \right),
\een
with $\f = {-1 \over N+1} \q$.
The assumption that only $i=1$ fermions are massive
is natural when we consider the fact that 
top quark is much massive than other quarks in real world. 
As the $U(1)$ current is bosonized as $J=\bar{\J}_{ai} \g_+
\J_{ai} \rightarrow i{MN \over 2 \p} \pp \q$, 
U(1) current interaction
is bosonized to be ${1 \over 8 \p}{MN^2 \over N+1} \pp \q \pb \q$ and shifts the
coefficent of the kinetic term for $\q$ in the action. This gives
the kinetic term for $\f$ in ${M \over 8 \p} {\Tr} l^{-1}
\pp l l^{-1} \pb l$ of eq.(15) to be
$-{1 \over 8 \p} {MN \over N+1} \pp \q \pb \q$.
As this term does
not contribute to the potential in bosonic picture, the soliton
configuration originated from the degeneracy of vacuum is
maintained. In fermionic picture
the current interaction term contributes to scattering process between
solitons by creating or annihilating fermion-antifermion pairs such that 
they dress the solitons. Note the
current interaction becomes negligible when we take large-N or large-M limit
in the fermionic picture, while the relative contribution to kinetic
term remains finite in bosonic picture.
The second term in the action is originated from the mass term 
and is explicitely $e^{-i(N+1) \f} (g_N) ^{-1} _{11} + e^{i(N+1) \f} 
(g_N) _{11}$ which coincide with the result from eq.(14).
So 2D-QCD with a massive fermion is bosonized to be an 
integrable theory deformed by the U(1) current interaction term.

\section{Integrability, solitons and conservation laws}
The action in eq.(15), especially the mass term 
bilinear in $l$, permits the equation of motion
to be expressed in a zero curvature form, leading to the integrable theory.\cite{shin1}\cite{shin2}
To see this, we first apply the variation on $l$ to obtain the equation of motion; 
\be
\{-\pb(l^{-1} \pp l) -m^2 [p,l^{-1} p l] \} l^{-1} \d l =0.
\ee
Using the simple relation
\be
\pp( l^{-1} p l) + [l^{-1} \pp l, l^{-1} p l] =0
\ee
eq.(17) can be recasted in a zero curvature form;
\be
[\pp + l^{-1} \pp l + \l p, \pb - {m^2 \over \l} l^{-1} p l] =0.
\ee

The integrability of the theory introduces the following Backlund
transformation[BT]\cite{shin4};
\ben
l^{-1} \pp l -{\tilde l}^{-1} \pp {\tilde l} + m^2 \h [ l^{-1}p {\tilde l},
 p]&=&0 \nonumber \\
\h \pb(l^{-1} p {\tilde l}) + l^{-1} p l - {\tilde l}^{-1}p {\tilde l}&=& 0
\een
where $\h$ is a BT parameter characterizing the solitons.
The Backlund transformation offers us the ability to calculate the 1-soliton
solution of the theory starting from the vacuum solution, for which $\tilde{l}
=1$, and multi-soliton solutions using non-abelian superposition rules\cite{shin4}.
When we parametrize $l$ for 1-soliton as
\be
l = \left( \begin{array}{rrrrr} e^{i \vf}& & & & \\ & 1 & & & \\
 & & \cdots & & \\ & & & 1 & \\ & & & & e^{-i \vf} \end{array} \right),
\ee
the BT becomes 
\ben
\pp \vf -2 m^2 \h \sin  \vf = 0 \nonumber \\
\h \pb \vf -2 \sin \vf = 0.
\een
The one  soliton solution that can be obtained from eq.(22) is the 
well-known sine-Gordon soliton which in the static case is\cite{shin3}
\be
\vf (x)=2 \tan^{-1} e^{4 m (x-x_0)}.
\ee
This solution interpolates two different vacuua of the theory, 
i.e. $ \vf(x \rightarrow -\infty)
=0,\ \vf(x \rightarrow \infty) = \p$, which are given
by the minimum of potential $-{M m^2 \over 2 \p}{\Tr} l^{-1} p l p 
=-{M m^2 \over \p} \cos 2 \vf$, i.e. $\vf = n \p$.

Another important aspect of the integrable theory is it has infinitely
many conserved quantities.\cite{shin2}
The interpretation of the zero curvature equation (19)
as a compatibility condition of two linearized equations permits us
to write the conserved currents in an iterative form.\cite{shin2}
Let us decompose the $(N+1) \times (N+1)$ matrix $l^{-1} \pp l $
and $l^{-1} p l$ as following;
\be
l^{-1} \pp l = \left( \begin{array}{ccccc} a+d& &c& &0\\ & & & & \\
-c^\dagger& &e& &0\\ & & & & \\
0& &0& & a-d \end{array} \right),\ \ 
l^{-1} p l = \left( \begin{array}{ccccc} 0& &0& &r-is\\ & & & & \\
0& &0& &q^\dagger\\ & & & & \\
r+is& &q& & 0 \end{array} \right)
\ee
where  $c, q$ are $1 \times (N-1)$ matrices
while $e$ is an $(N-1) \times (N-1)$ matrix with property $e = -e^\dagger$. 
$a, d$ are pure imaginary numbers while $r, s$ are real.
Then the  conserved currents are calculated iteratively with 
these matrix elements by following formulae;
\ben
 & &\a_i = {-1 \over m^2} \{ ( \pp -a+e) \a_{i-1} -{1 \over 2} c^\dagger
(\j_{i-1} + \b_{i-1}) \} \nonumber \\
 & &\b_i = {-1 \over 2 m^2} ( \pp \b_{i-1} + c \a_{i-1} + d \j_{i-1}) \nonumber \\
 & &J_i = \pp \j_i ={1 \over m^2}\{ (c \pp -ac +ce +{1 \over 2} dc) \a_{i-1}
+{1 \over 2} (d \pp -c c^\dagger ) \b_{i-1} -{1 \over 2} (c c^\dagger -d^2) 
\j_{i-1} \} \nonumber \\
 & &\bar{J}_i = \pb \j_i = -q \a_{i-1} -is \b_{i-1} -r \j_{i-1}.
\een
Now stating with $\a_0 = \b_0 =0, \j_0 =1$,  all higher 
currents satisfying $\pb J_i = \pp \bar{J}_i$ can be calculated.
For i=1, the conservation law becomes
\be
{1 \over 2 m^2}\pb (cc^\dagger -d^2) = \pb r.
\ee
This is just the conservation of energy.
Higher order conserved currents in general become non-local. For example,
\ben
 & &J_2 ={1 \over 2 m^4} \{ c(\pp -a+e+{d \over 2} ) c^\dagger -{1 \over 2} (d \pp 
-c c^\dagger) d \} + J_1 \j_1 \nonumber \\
 & &\bar{J}_2 ={-1 \over 2 m^2} (q c^\dagger -isd) \} + \bar{J}_1 \j_1.
\een
This however can be made local if we subtract non-local terms containing $\j_1$
from currents. Note that the non-local terms themselves satisfy the conservation
law, $\pb (J_1 \j_1) = \pp (\bar{J}_1 \j_1)$. Similarly the third-order currents
after subtracting non-local terms containing $\j_1, \j_2$ are
\ben
 & &J_3 = {-1 \over 2 m^6} c ( \pp-a+e+{d \over 2} )^2 c^\dagger
-{1 \over 4 m^6} (c c^\dagger -{d^2 \over 2} ) (c c^\dagger -d^2)
+{1 \over 8 m^6} (d \pp - c c^\dagger )(\pp d - c c^\dagger) \nonumber \\
 & &\bar{J}_3 = {1 \over 2m^4} q ( \pp -a+e+{1 \over 2} d) c^\dagger
-{1 \over 4 m^4} is ( \pp d -c c^\dagger).
\een
These conservation laws can be directly checked using the equation of motions
(17),(18) which can be expressed in a new form as
\ben
 & &\pb a = \pb e = 0,\ \ \pb c + m^2 q =0,\ \ \pb d +2ims =0 \nonumber \\
 & &\pp q +(a-d) q -(r+is) c -q e=0 \nonumber \\
 & &\pp r-2ids +{1 \over 2}(c p^\dagger + p c^\dagger) =0 \nonumber \\
 & &\pp s +2idr +{i \over 2} (c p^\dagger -p c^\dagger)=0.
\een
These infinitely many conservation laws guarantee the shape-preserving property
of solitons after collisions and will be helpful in fixing the scattering
amplitudes.\cite{abd1} 2D-QCD is an almost integrable system
and the difference from the
exactly integrable system is the change of the coefficient of U(1) kinetic
term. We speculate the baryons after collision can have zero-soliton quantum
fluctuations around them, making a slight deformation of solitons.

\section{Discussions}
Finally we point out the different nature of quantum 
fluctuation around solitons of present theory from the conventional 
one.\cite{frish5}
Usually the quantum fluctuation is described by the matrix $l(x,t) = A(t) l_0(x) A^{-1} (t)$ 
\cite{frish1} where
\be
A = \left( \begin{array}{ccccc} z_1 & y_2 & \cdots & y_N & 0 \\ 
z_2 & & & & 0 \\
 & & a_{ij} & & \\ z_N & & & & \\ 0 & & 0 & & \z \end{array} \right).
\ee
This form describes finite-energy configuation around soliton when 
the soliton has the property $l_0(x) \rightarrow 1$ as $ 
x \rightarrow \pm \infty$. 
In the present theory $l_0(x) \not \rightarrow 1$ when $ x \rightarrow \infty$, and
$l(x,t)$ does not correspond to finite-energy configuration.
Indeed the mass term is
\be
{\Tr} l^{-1} p l p - {\Tr} l_0 ^{-1} p l_0 p 
=2(\cos 2 \vf -\cos \vf ) (z_1 z_1 ^* -1).
\ee
As the soliton  $\vf \rightarrow \p$ when $x \rightarrow \infty$, 
the energy of new configuration becomes infinite except the case $
|z_1| \rightarrow 1$ as $x \rightarrow \infty$.
Similarly the term $[A^{-1} \dot{A}, l_0][A^{-1} \dot{A}, l_0 ^\dagger]$ 
contained in $S_0 [A l_0 A^{-1} ] - S_0 [l_0 ]$
becomes
\be
4(1-\cos \vf )\{ \dot{z}_i ^* \dot{z}_i + \dot{\z} ^* \dot{\z}
 +(z_i ^* \dot{z}_i )^2 +(\z ^* \dot{\z} )^2 \},
\ee
which also becomes infinite except the case $|z_1| \rightarrow 1$.
But when we take $|z_1| = 1$ without dependence on $x$,
it only results $l(x,t) = l_0(x)$. So x-dependence of
$z_1$ and $A$ is necessary to make a finite-energy configuration, 
resulting a pulsating solution. The detailed study of this configuration is 
not performed yet and we defer it for future analysis.  

Recently there appears extensive development called the decoupled bosonization.\cite{abd1}-\cite{abd5}
It bosonize 2D-QCD 
by rewriting the theory in terms of gauge invariant fields and describes
massless fermions in terms of positive and negative level WZW models, ghosts and
massive bosonic excitations.
Like our formalism it also has led to interesting insights into the characterestics of the model,
such as its integrability, degeneracy of the vacuum, and higher symmetry
algebras.
And its integrability condition is
valid for the quantum 
theory as well. But this model displays a complicated set of constraints
and the expression for the integrability is non-local.
So it seems difficult to find explicitly the classical soliton solutions and
to analyze the mesonic spectrum and their physical properties
using the method of Skyrme model.  In addition the integrability condition
does not survive for massive case in this approach.\cite{abd4}
It can be thought that our formulation, specially
the method treating the mass term, could find interesting application
in the decoupled formulation. One interesting common feature shared by
two formulations is the so-called quasi-integrability, 
that is particle pair production
is not entirely suppressed.\cite{abd5} It seems worthwhile to make a
detailed analysis of two formulations simultaneously and make a
simple and exact bosonization formulation taking merit of each formalism.

\vglue .3in 
{\bf ACKNOWLEDGEMENT}
\vglue .2in
We would like to thank to Professor Q-Han Park for discussion. 
This work was supported in part by the program of Basic Science Research, 
Ministry of Education BSRI-96-2442, and by Korea Science and Engineering 
Foundation through CTP/SNU.
\vglue .2in

\end{document}